# Association Of Unidentified, Low Latitude EGRET Sources With Supernova Remnants

Steven J. Sturner[1,2] and Charles D. Dermer[1]

[1] E. O. Hulburt Center for Space Research, Code 7653, Naval Research Laboratory, Washington, DC 20375-5352, USA
[2] National Research Council Research Associate

**Abstract.** We propose that some of the unidentified Galactic plane ($-10° < b < 10°$) sources listed in the first EGRET catalog (Fichtel et al. 1994) are associated with nearby supernova remnants. The probability that the association of strong EGRET sources with supernova remnants is due to chance alignment is found to be $\lesssim 1/2200$. Three of the most convincing associations are radio-bright, nearby shell remnants and a fourth is associated with a plerion. We examine possible gamma-ray production mechanisms and conclude that pion production and decay from high-energy protons interacting with the remnant gas could produce the gamma rays. The energetics of this process are in accord with theories of cosmic ray production through supernova shock acceleration. The identification of $\pi^o$-decay features in the gamma-ray spectra of supernova remnants would support this interpretation.

## 1. Introduction

The identified high-energy gamma-ray ($\gtrsim 100$ MeV) sources observed with the EGRET experiment on the Compton Gamma-Ray Observatory have fallen into five general categories: solar flares, pulsars, gamma-ray bursts, normal galaxies, and active galactic nuclei. In addition, the first EGRET source catalog (Fichtel et al. 1994) lists 37 unidentified Galactic plane sources, of which 10 are high-confidence detections ($> 6\sigma$) and 27 are marginal detections (between 5 and $6\sigma$). We have compared the locations of these 37 sources with the locations of the 182 supernova remnants (SNRs) listed in the catalog of Green (1994). For simplicity, the radii of elliptical SNRs were taken to be the average of the semi-major and semi-minor axes of the remnant. We found that of the 37 unidentified high-energy gamma-ray sources, 4 of the high confidence sources and 9 of the marginal detections have SNRs whose radii overlap the 95% confidence EGRET gamma-ray error circle (one of the marginal detections has two possible associations; see Table 1). Two of the gamma-ray sources have SNRs within $8'$ of the centroid of the EGRET error circle. One of these two sources is at a Galactic longitude of 189° where there are few SNRs and the possibility of chance coincidence is low.

## 2. Statistical Test

We developed a test to determine if the associations between the 37 EGRET sources and the 182 SNRs are chance alignments. We calculate the angular distance, as measured from the center of the EGRET error circle to the center of the remnant, from each gamma-ray source to the closest three SNRs, $\theta_1, \theta_2$, and $\theta_3$. Assuming that the nearest neighbors are randomly distributed about the EGRET source, we then use these distances to define two parameters, $a_1 \equiv (\theta_1^2/\theta_2^2)$ and $a_2 \equiv (\theta_2^2 - \theta_1^2)/(\theta_3^2 - \theta_1^2)$. The quantity $a_1$ is the ratio of the areas of circles with radii $\theta_1$ and $\theta_2$ centered at the location of the EGRET source. The quantity $a_2$ is the ratio of the areas of annuli with inner radius $\theta_1$ and outer radii $\theta_2$ and $\theta_3$, respectively, centered on the EGRET source. If the SNRs are not correlated with the EGRET sources, then $\langle a_1 \rangle$, the average value of $a_1$, should equal 0.5 within statistical errors. If the unidentified EGRET sources are, however, associated with SNRs, then $\langle a_1 \rangle$ should be $< 0.5$. The average value of $a_2$, $\langle a_2 \rangle$, should be 0.5 if our assumption regarding the distribution of nearest neighbors is correct.

Another way of testing the assumption that the distribution of the nearest-neighbor SNRs about an unidentified EGRET source is two-dimensional is to compute the average of the sines of the angles between the directions from the EGRET source to the nearest and second nearest (second and third nearest) SNRs, $\langle \sin\psi_{1(2)}\rangle$. If the SNR distribution is two-dimensional, $\langle \sin\psi \rangle = 2/\pi = 0.637$. If the distribution is linear, the average should be much less than 0.637. We find that for the 10 high confidence sources $\langle \sin\psi_{1(2)} \rangle = 0.69(0.47) \pm 0.10$ where the $1\sigma$ error is given by $(0.5 - 4/\pi^2)^{1/2}/\sqrt{N}$, where N is the number of sources. For the 27 marginal sources, $\langle \sin\psi_{1(2)} \rangle = 0.53(0.64) \pm 0.06$ and for the entire sample $\langle \sin\psi_{1(2)} \rangle = 0.58(0.60) \pm 0.05$. Thus our assumption appears valid.

**Table 1.** Unidentified EGRET Sources With Possible SNR Associations

| EGRET Source | SNR | $\theta_1$ (')[a] | $D_{max}$ (')[b] | $\theta_1/D_{max}$ | Type[c] | Radio Flux (Jy) |
|---|---|---|---|---|---|---|
| GRO J0542+26 | G 180.0-1.7 (S147) | 116.6 | 248.0 | 0.47 | S | 65 |
| GRO J0617+22 | G 189.1+3.0 (IC 443) | 6.7 | 43.5 | 0.15 | S | 160 |
| GRO J1110-60 | G 291.0-0.1 (MSH 11-62) | 7.8 | 56.0 | 0.14 | F | 16 |
| GRO J2019+40 | G 78.2+2.1 ($\gamma$ Cygni) | 27.7 | 48.0 | 0.58 | S | 340 |
| GRO J0635+05 | G 205.5+0.5 (Monoceros) | 81.6 | 148.0 | 0.55 | S | 160 |
| GRO J0823-46 | G 263.9-3.3 (Vela) | 87.7 | 127.5 | 0.69 | C | 1750 |
| GRO J1416-61 | G 312.4-0.4 | 11.9 | 49.8 | 0.24 | S | 44 |
| GRO J1443-60 | G 316.3+0.0 (MSH 14-57) | 35.4 | 39.0 | 0.91 | S | 24 |
| GRO J1758-23 | G 6.4-0.1 (W28) | 26.8 | 57.0 | 0.47 | C | 310 |
| GRO J1823-12 | G 18.8+0.3 (Kes 67) | 13.0 | 36.8 | 0.35 | S | 27 |
| GRO J1842-02 | G 30.7+1.0 | 42.3 | 50.5 | 0.84 | S | 6 |
| GRO J1853+01 | G 34.7-0.4 (W44) | 30.7 | 42.5 | 0.72 | S | 230 |
| GRO J1904+06 | G 40.5-0.5 | 36.7 | 63.0 | 0.58 | S | 11 |
|  | G 41.1-0.3 (3C397) | 46.1 | 53.8 | 0.86 | S | 22 |

[a] Angular distance from center of EGRET error circle to center of associated remnant.
[b] Sum of the EGRET error circle radius plus the radius of the associated remnant.
[c] S=Shell, C=Composite, F=Filled

We find that given the entire sample of 37 unidentified high-energy gamma-ray sources, $\langle a_1 \rangle = 0.41 \pm 0.05$ and $\langle a_2 \rangle = 0.45 \pm 0.05$, where the errors are statistical, i.e. $\sigma = 1/\sqrt{12N}$ where $N$ is the number of sources. Thus there is $\sim 1.8\sigma$ deviation of $\langle a_1 \rangle$ from 0.5 while $\langle a_2 \rangle$ deviates by only $1\sigma$ from 0.5. This weakly suggests that the distribution of SNRs closest to the gamma-ray sources is not completely random while supporting the assumption that the distribution of second closest remnants is random. When only the 10 high confidence sources are studied, however, the values of $\langle a_1 \rangle$ and $\langle a_2 \rangle$ are $0.21 \pm 0.09$ and $0.53 \pm 0.09$, respectively. In this case, there is $\sim 3.2\sigma$ deviation from 0.5 for $\langle a_1 \rangle$ while $\langle a_2 \rangle$ is within $1\sigma$ of 0.5, suggesting only a small chance of coincidental alignment with the closest remnants. Computer simulations show that the probability of having $\langle a_1 \rangle \lesssim 0.21$ is $\lesssim 1/2200$.

When considering only the marginal gamma-ray source detections, $\langle a_1 \rangle = 0.48 \pm 0.06$ and $\langle a_2 \rangle = 0.43 \pm 0.06$, both roughly consistent with 0.5. The lack of statistically significant association between the weak EGRET-source and supernova-remnant populations is in part because most marginal EGRET detections are located toward the galactic centre region where there is a larger density of SNRs, and the relatively large size of the EGRET error circle reduces the probability of association. We show histograms of the values of $a_1$ for the high-confidence, marginal, and total set of EGRET sources in Figure 1. Note that all but one of the high-confidence sources has a value for $a_1 < 0.5$. We would like to note that Esposito et al.(1994) have independently found a statistically significant correlation between the unidentified EGRET sources and SNRs using a different statistical test.

## 3. Properties of the Associated Supernova Remnants

We now consider the energetics of the SNRs associated with the high confidence gamma-ray detections. Of these, three are radio-bright, nearby shell-type remnants (Green 1994), namely $\gamma$ Cygni (G 78.2+2.1), IC 443 (G 189.1+3.0), and S147 (G 180.0-1.7), and the fourth, MSH 11-62 (G 291.0-0.1), is a plerion (Weiler & Sramek 1988). The distance to IC 443 is between 0.7 and 2.0 kpc (Green 1994), and Petre et al. (1988) have estimated its age at 2800-3400 years. $\gamma$ Cygni has an estimated distance of 1.5 kpc (Green 1989; Landecker, Roger, & Higgs 1980). An age for $\gamma$ Cygni can be estimated assuming that this remnant is in the Sedov phase using the equation (Kassim, Hertz, & Weiler 1993)

$$\tau \approx 0.1 \, (d \, \theta)^{2.5} \left( \frac{n_o}{\epsilon_{51}} \right)^{0.5} \text{ years,} \qquad (1)$$

where $d$ is the distance to the remnant in kpc, $\theta$ is the angular diameter of the remnant in arcminutes, $n_o$ is the density of the interstellar medium surrounding the remnant in cm$^{-3}$, and $\epsilon_{51}$ is the supernova explosion energy in units of $10^{51}$ ergs. The ISM density $n_o$ typically has values between 0.1 and 1.0 cm$^{-3}$. For simplicity we choose 2 shell remnants. The age of this remnant along with the large 95% confidence error circle (158′) and the large size of the remnant make chance alignment a distinct possibility, although it is located in a region with few other remnants. MSH 11-62 is a plerion with an age of 3000 years and a distance of $\sim 3.5$ kpc (Roger et al. 1986). Therefore its gamma rays could originate from high-energy particles emitted by the pulsar which is presumably responsible for powering this remnant.

We have ordered Green's (1994) catalog of SNRs according to radio flux. $\gamma$ Cygni and IC 443 are the fifth (340 Jy) and eleventh (160 Jy) brightest SNRs at 1 GHz, respectively. S147 and MSH 11-62 are significantly dimmer in the radio, and are numbers 25 (65 Jy) and 74 (16 Jy) on the list. When compared with only the flatter spectrum (spectral index $\lesssim 0.5$) shell remnants, $\gamma$ Cygni and IC 443 are the second and seventh brightest. The radio bright SNRs that are not associated with EGRET sources tend to be either farther away than $\gamma$ Cygni and IC 443, such as Cas A and W28 at 2.8 and 3.5-4.0 kpc (Green 1994), respectively, or are much older, such as the Lupus Loop at $\sim 50,000$ years (Leahy, Nousek, & Hamilton 1991).

## 4. Possible Gamma-Ray Production Mechanisms

Gamma-ray emission from MSH 11-62 can be readily understood as either pulsed emission from an associated radio-quiet pulsar with unknown period or as unpulsed emission from a pulsar-energized nebula (plerion) such as the Crab Nebula. Detection of a point source in X-rays by *ROSAT* or *ASCA* would give evidence for a pulsar source for the emission and could permit a search for the pulsar's period. Gamma rays from shell-type remnants such as $\gamma$ Cygni and IC 443 (and possibly S147) may involve other production mechanisms, although searches for radio-quiet pulsars associated with these sources should also be pursued. Let us therefore consider the possible origin of the gamma rays from the shell-type SNRs. It is not likely that the high-energy radiation originates from Compton upscattering of a photon field by the synchrotron-emitting electrons. For example, IC 443 at an assumed distance of 1.5 kpc, emits $\approx 10^{33}$ ergs s$^{-1}$ in the radio (Green 1994; Mufson et al. 1986), $\approx 10^{37}$ ergs s$^{-1}$ in the IR (Mufson et al. 1986), and $\approx 10^{35}$ ergs s$^{-1}$ in X-rays (Petre et al. 1988; Mufson et al. 1986). If our proposed association is correct, IC 443 emits $\approx 5 \times 10^{34}$ ergs s$^{-1}$ in $> 100$ MeV gamma rays (Fichtel et al. 1994). The relationship between the gamma-ray luminosity $L_\gamma$ and the radio luminosity $L_r$ from Compton upscattering and synchrotron radiation, respectively, can be shown to be

$$\frac{L_\gamma}{L_r} \approx 50 \approx \frac{L_{ph}}{4\pi r^2 (B^2/8\pi)c}, \qquad (2)$$

where $L_{ph}$ is the luminosity of one of the lower-energy photon fields, $B$ is the mean magnetic field strength in $n_o/\epsilon_{51} = 0.5$. Thus $\tau \approx 0.07 \, (d \, \theta)^{2.5}$ years. For $\gamma$ Cygni, $\tau \approx 5400$ years given its 60′ diameter (Green 1994). For comparison, Equation (1) gives an age of 13,000 years for the Vela supernova remnant which is in good agreement with the 11,000 year spin-down age for the Vela pulsar. On the other hand, S147 is a nearby, old shell-type remnant. Sauvageot, Ballet, & Rothenflug (1990) give its distance as 1.1 − 1.7 kpc. Given its 180′ diameter, Equation (1) yields an age of $\sim$70,000 years, much older than the other the remnant shell [typically of order $10^{-5}$ Gauss (Lozinskaya 1992)], and $r$ is the remnant radius (the remnant thickness does not enter). Equation (2) implies that $B(10^{-5}\text{ Gauss}) \approx 0.01(L_{ph}/10^{37}\text{ergs s}^{-1})^{1/2}/(r/10\text{pc})$. Thus, unless the remnant magnetic field is extremely weak, there are not enough synchrotron-emitting electrons to produce the observed gamma rays by upscattering the known radiation fields or, for that matter, the microwave background or starlight radiation fields.

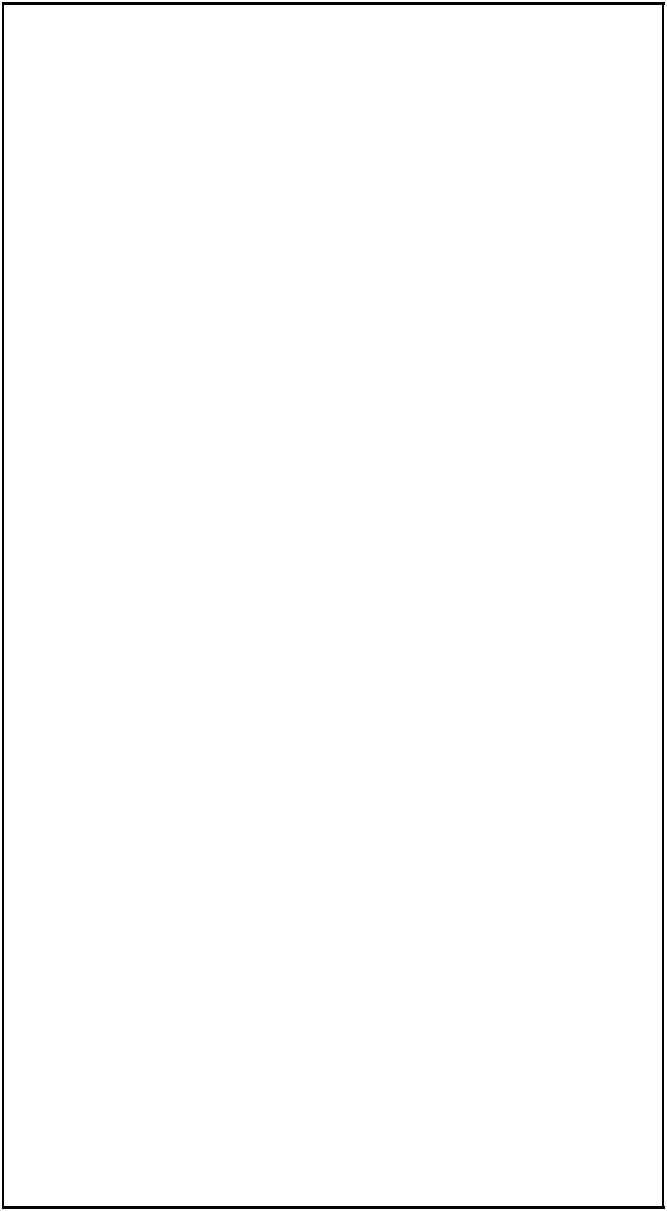

**Fig. 1.** A histogram of the number of EGRET unidentified sources with different values of the parameter $a_1 = \theta_1^2/\theta_2^2$, where $\theta_1$ and $\theta_2$ are the angular distances to the closest and second closest supernova remnants, respectively. Note the large fraction of high confidence EGRET sources which have values of $a_1 < 0.1$, indicating likely associations.

The relativistic electrons which produce the observed radio synchrotron radiation will also produce gamma-rays through nonthermal bremsstrahlung. The nonthermal electron spectrum in IC 443 can be estimated by comparing the derived synchrotron photon luminosity with its observed radio luminosity. The synchrotron or bremsstrahlung luminosity can be written as

$$L = m_e c^2 \int_1^{\gamma_o} N_e(\gamma_e)\dot\gamma d\gamma_e, \quad (3)$$

where, for synchrotron radiation, $\dot\gamma = \dot\gamma_{syn} \cong (4/3)c\sigma_T\gamma_e^2(B^2/8\pi m_e c^2)$ and $N_e(\gamma_e) = N_e^o\gamma_e^{-q}$ is the electron density per unit $\gamma_e$. In order to produce radio photons with frequencies near 10 GHz, $\gamma_o \gtrsim 1.9 \times 10^4 B_{-5}^{-0.5}$, where $B_{-5}$ is the magnetic field in units of $10^{-5}$ Gauss. The spectral index of the electron population, $q$, must $\approx 1.72$ in order to produce the 0.36 radio spectral index observed for IC 443 (Green 1994). Setting Equation (3) equal to $10^{33}L_{33}$ ergs s$^{-1}$, we find $N_e^o \approx 4 \times 10^{52} B_{-5}^{-1.36}L_{33}$. For bremsstrahlung radiation, $\dot\gamma = \dot\gamma_{brem} \cong (1/3)c\sigma_T\alpha_f n_H\gamma_e$, and $n_H = 10n_1$ cm$^{-3}$ is the mean number density of hydrogen atoms in the remnant. The derived bremsstrahlung luminosity $\approx 9 \times 10^{32} n_1 B_{-5}^{-1.5} L_{33}$ ergs s$^{-1}$, which implies $B(10^{-5}\text{ Gauss}) \approx 0.07(n_1 L_{33})^{0.67}$ given the observed gamma-ray luminosity of IC 443.

It was suggested long ago that if SNRs are sites of cosmic-ray acceleration, then $\pi^o$-decay gamma-rays will be produced as a result of collisions of the cosmic rays with the interstellar medium (Pinkau 1970). The luminosity in secondary $\pi^o$s is

$$L_{\pi^o} \cong cn_H \int_{\gamma_{p,thresh}\approx 2}^\infty N_p(\gamma_p)\sigma_{pp}(\gamma_p)\langle E(\gamma_p)\rangle d\gamma_p, \quad (4)$$

where $N_p(\gamma_p) = N_p^o\gamma_p^{-2.2}$ is the number of protons per unit $\gamma_p$ suggested by shock acceleration models (Gaisser 1990), $\sigma_{pp} \approx 3 \times 10^{-26}$cm$^{-2}$ (Dermer 1986) is the proton-proton interaction cross section, and $\langle E(\gamma_p)\rangle \cong 1/3(\gamma_p m_p c^2/2)$ is the average amount of energy per proton-proton interaction converted to $\pi^o$-decay gamma rays. By setting Equation (4) equal to the gamma-ray luminosity of the EGRET source possibly associated with IC 443, $\approx 5\times 10^{34}$ ergs s$^{-1}$, we find $N_p^o \approx 5 \times 10^{51} n_1^{-1}$ protons. Thus the amount of energy in nonthermal protons required to produce the observed gamma-ray emission is $\approx 4 \times 10^{49} n_1^{-1}$ ergs. This can be compared with $E_{cr}$, the amount of energy per supernova that is converted to nonthermal protons in order to produce the observed cosmic ray energy density $\Psi_{cr}$, which we write as $E_{cr} \cong \Psi_{cr}V_{gal}\tau_{sn}/\tau_{gal}$. Here $\Psi_{cr} = 10^{-12}\Psi_{-12}$ ergs cm$^{-3}$ (Gaisser 1990), $V_{gal} = 10^{67} V_{67}$ cm$^3$ is the volume of the galaxy, $\tau_{sn} = 10^9\ t_{sn,9}$ seconds is the time between supernovae in the galaxy, $\tau_{gal} = 3 \times 10^{14}\ t_7$ seconds is the proton confinement time in our galaxy, and $t_7$ is the confinement time in units of $10^7$ years. Thus $E_{cr} \approx 3 \times 10^{49}\ \Psi_{-12}V_{67}t_{sn,9}/t_7$ ergs, similar to that needed to produce the observed gamma-ray flux. Aharonian, Drury, and Völk (1994) have also found that gamma rays produced by the decay of $\pi^o$s could be detectable with current instruments, particularly if the remnant is adjacent to a molecular cloud as is evidently the case for $\gamma$ Cygni and IC 443 (Pollock 1985; Asaoka & Aschenbach 1994) Observations of pion-decay features at gamma-ray energies from SNRs with the telescopes on the *Compton Observatory* would support the $\pi^o$ origin of the gamma-ray emission and be in accord with the scenario that cosmic rays are accelerated by supernova shocks.

*Acknowledgements.* We thank N. Kassim, K. Weiler, J. Skibo, and S. Lundgren for suggestions and discussions. The work of S.S was supported by NASA grant 93-085.

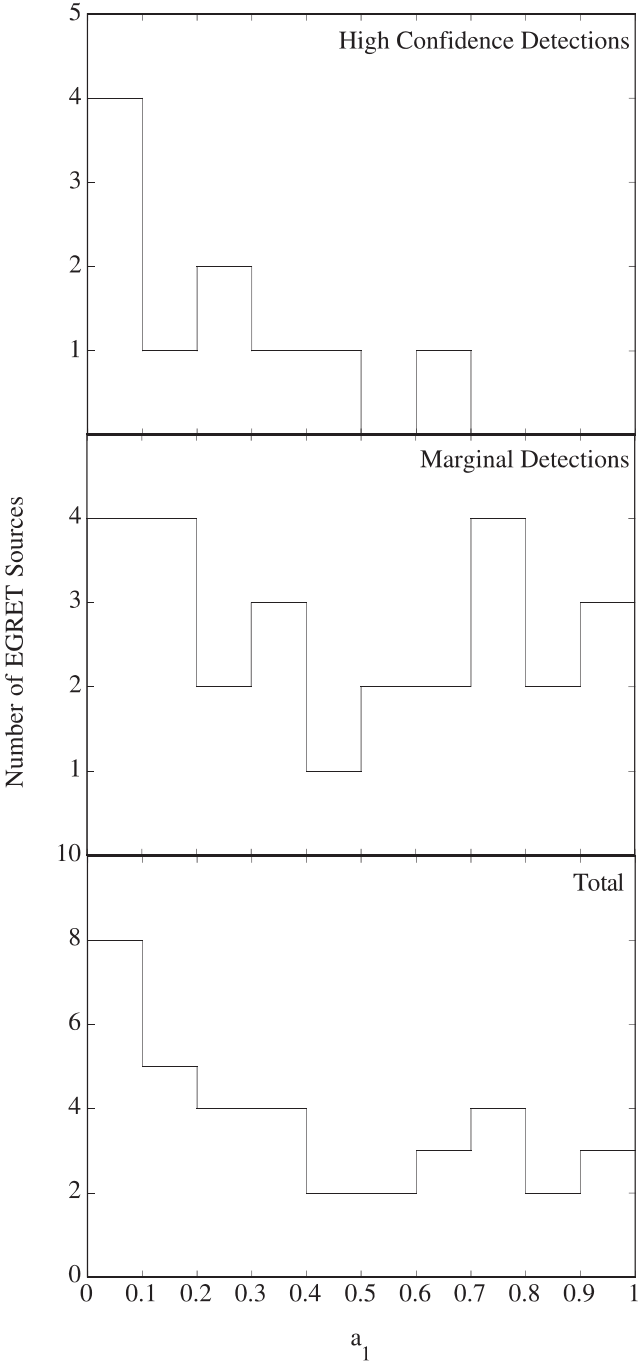

Figure 1